\documentclass{aa}  

\usepackage{graphicx}
\usepackage{txfonts}
\usepackage{lipsum}
\usepackage{subcaption}         
                                
\usepackage{lscape}            
                               
\usepackage{placeins}           
                               
\usepackage{bm}
\usepackage{amsmath}    
\usepackage{amssymb}    
\usepackage[colorlinks=true,allcolors=blue,linkcolor=red]{hyperref}

\usepackage{xcolor}

\begin{document}

\title{{Characterization and formation of the Mg\,\textsc{i} 12.32 $\mathrm{\mu}$m line in the quiet Sun and sunspot}}
\titlerunning{Characterization and formation of the Mg\,\textsc{i} 12.32 $\mathrm{\mu}$m line in the quiet Sun and sunspot}
\authorrunning{Wu et al.}

\author{Yuchuan Wu\inst{1}\fnmsep\inst{2}, Wenxian Li\inst{1}, Xianyong Bai\inst{1}\fnmsep\inst{2}, Feng Chen\inst{3}\fnmsep\inst{4}, Hao Li\inst{5} \and Yuanyong Deng\inst{1}\fnmsep\inst{2}}

\institute{State Key Laboratory of Solar Activity and Space Weather, National Astronomical Observatories, Chinese Academy of Sciences, Beijing, China \email{wuyc@bao.ac.cn; wxli@nao.cas.cn} \and 
School of Astronomy and Space Science, University of Chinese Academy of Sciences, Beijing, China\and
School of Astronomy and Space Science, Nanjing University, Nanjing, China\and
Key Laboratory of Modern Astronomy and Astrophysics (Nanjing University), Ministry of Education, Nanjing, China\and
State Key Laboratory of Solar Activity and Space Weather, National Space Science Center, Chinese Academy of Sciences
 }
 
  \abstract
   {The Mg\,\textsc{i} 12.32 $\rm{\mu}$m line is highly sensitive to magnetic fields due to its long wavelength, making it a promising tool for precise solar-magnetic-field measurements. The formation of this line is significantly influenced by nonlocal thermodynamic equilibrium (NLTE) effects.}
   {Previous studies have shown that the Mg\,\textsc{i} 12.32 $\rm{\mu}$m line exhibits different behaviors in various regions of the Sun. This study focuses on the peak intensity of the Mg\,\textsc{i} 12.32 $\rm{\mu}$m line to analyze its relationship with the physical parameters of the solar atmosphere and its formation mechanism.} 
   {We employed the Rybicki-Hummer (RH) 1.5D radiative transfer code to synthesize the Stokes profiles of the Mg\,\textsc{i} 12.32 $\rm{\mu}$m line based on a three-dimensional solar atmospheric model of a sunspot and its surrounding quiet {Sun}. 
   By computing $\overline{R}_{x_i}\Delta x_i$, where $\overline{R}_{x_i}$ is the average response function and $\Delta x_i$ is the difference in physical parameters between the two models being compared, we identified the atmospheric height and physical parameters that most significantly influence the normalized peak intensity in the quiet {Sun} and the active region, respectively.
   }
   {In analyzing the synthesized Stokes profiles, we found two key features: (1) in the quiet {Sun}, the normalized peak intensity is strong at the centers of the granules and weakens in the intergranular lanes; (2) in the sunspot umbra, the normalized peak intensity is generally weak, with only a few areas showing evident emission. Through the analysis of the response functions, we identified the causes of these differences. 
In the quiet Sun, the differences in normalized peak intensity are primarily attributed to temperature variations at {log $\tau_{\rm{500}}$; the logarithm of the continuum optical depth at $\lambda=500$ nm,} ranging from {$-$0.21 to 0.91} and from {$-$1.65 to $-$0.76}; as well as to temperature and density variations at {log $\tau_{\rm{500}}$} ranging from {$-$3.86 to $-$2.38}. In the sunspot umbra, the differences are mainly due to density variations at {log $\tau_{\rm{500}}$} ranging from {$-$0.96 to 1.26}. In addition, we discussed the mechanisms through which these physical parameters influence the normalized peak intensity.}
   
   {}


   \maketitle

\section{Introduction}

The Mg\,\textsc{i} 12.32 $\rm{\mu}$m line is an nonlocal thermodynamic equilibrium (NLTE) line that has significant potential for solar magnetic-field measurement due to its long wavelength and high sensitivity to magnetic fields \citep{1992A&A...253..567C,1983ApJ...269L..61B}. \citet{1981ApJ...247L..97M} first observed two spectral lines near 12 $\rm{\mu}$m in the solar spectrum, which are characterized by emission peaks and wide absorption troughs. \citet{1983ApJ...275L..11C} further identified them as transitions between high Rydberg states, i.e., $3s7h^{1,3}\rm{H}^o-3s6g^{1,3}\rm{G}^e$ (12.22 $\rm{\mu}$m) and $3s7i^{1,3}\rm{I}^e-3s6h^{1,3}\rm{H}^o$ (12.32 $\rm{\mu}$m), of neutral magnesium (Mg\,\textsc{i}). The Land\'{e} g-factor of the pair lines was determined to be unity \citep{1987PhyS...35..792C,1988A&A...191L...4L}. 

\begin{figure*}[h!]
\centering
\includegraphics[width=\hsize]{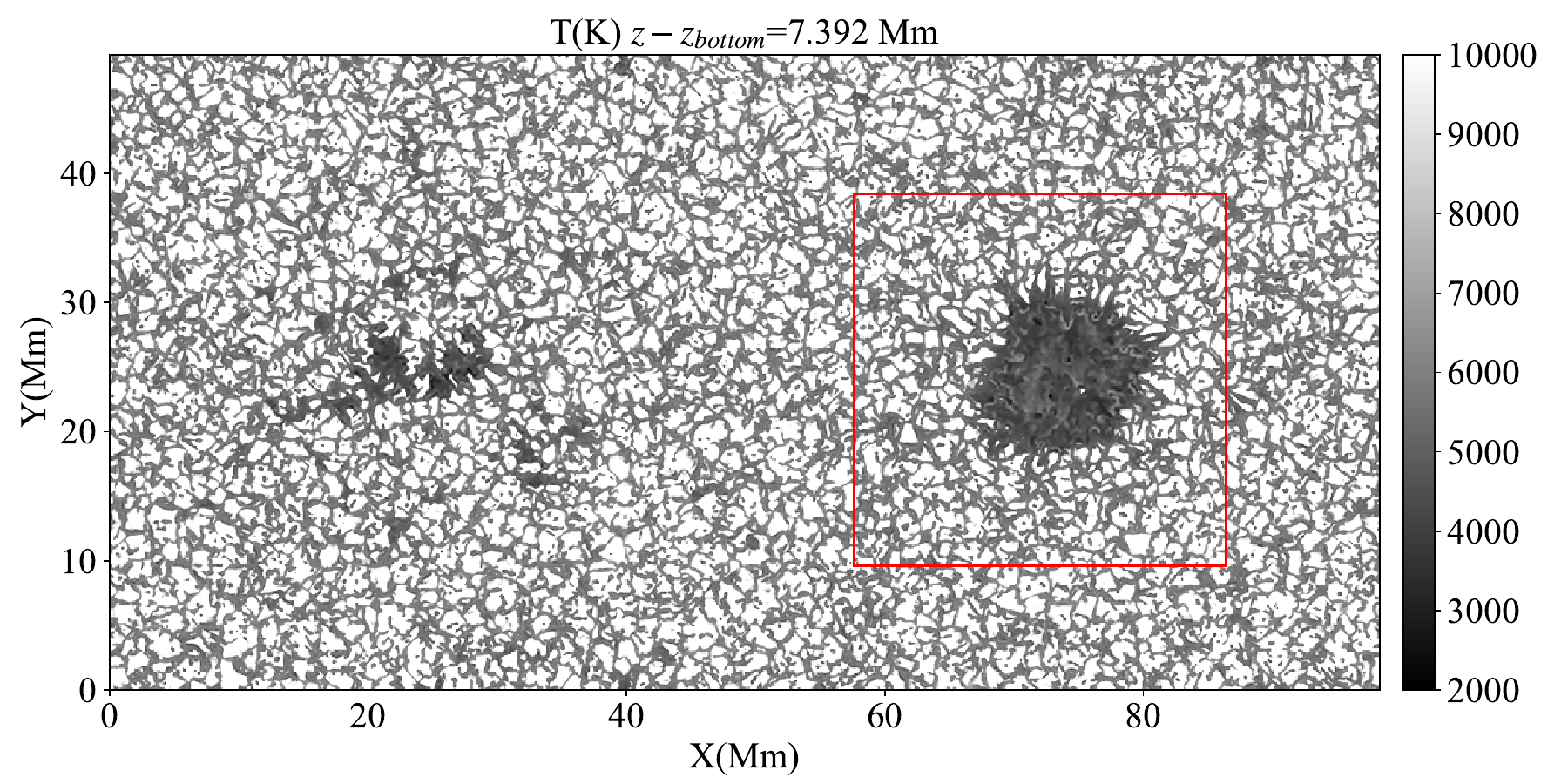}
\caption{ 
Temperature at 7.392 Mm above bottom boundary of model{, corresponding approximately to $\rm{log}\, \tau_{500}=0$ in the quiet Sun, where $\tau_{500}$ is the continuum optical depth at $\lambda=500$ nm.} The red rectangle delimits the region of interest for the forward synthesis shown in Fig.~\ref{fig2}.
}
\label{fig1}
\end{figure*}

To date, observations of the Mg\,\textsc{i} 12.32 $\rm{\mu}$m spectral line have mainly been obtained from the McMath-Pierce Telescope \citep{1964ApOpt...3.1337P}. 
Based on the observations from McMath-Pierce, studies have been conducted on the spectral characteristics of the Mg\,\textsc{i} 12.32 $\rm{\mu}$m line. 
For example, \citet{1983ApJ...269L..61B} reported the following features of the Mg\,\textsc{i} 12.32 $\rm{\mu}$m spectral line: (1) the Mg\,\textsc{i} 12.32 $\rm{\mu}$m line exhibits limb brightening, and the absorption troughs disappear near the solar limb;
(2) in the quiet {Sun}, sunspot penumbra, and plages, the Mg\,\textsc{i} 12.32 $\rm{\mu}$m spectral line exhibits emission, while it disappears in sunspot umbra;
(3) pronounced splitting, rising from the Zeeman effect, can be observed in the sunspot penumbra and plages.
\citet{1990ApJ...364L..49D} observed the Mg\,\textsc{i} 12.32 $\rm{\mu}$m emission line in the umbra during a solar flare. 
Due to its high sensitivity to magnetic fields, the Mg\,\textsc{i} 12.32 $\rm{\mu}$m line has been used to diagnose the solar magnetic field. 
By directly measuring the Zeeman splitting, \citet{1989ApJ...340..571Z} measured the magnetic field in various solar features, including plages, sunspot penumbrae, and the quiet {Sun}. 
\citet{1993ApJS...86..313H} utilized full Stokes profiles to measure parameters such as magnetic-field strength, inclination, azimuth, and magnetic-field filling factors. \citet{2007SoPh..241..213M} constructed the solar vector magnetogram using the Mg\,\textsc{i} 12.32 $\rm{\mu}$m spectral line.

In addition to observations, theoretical investigations of the formation of the Mg\,\textsc{i} 12.32\,$\rm{\mu}$m have been carried out. 
By employing an NLTE radiative-transfer calculation, \citet{1992A&A...253..567C} successfully reproduced the observed features of this line, including the emission peak and absorption trough, as well as the center-to-limb variation.
They confirmed the previous conclusion by \citet{1987A&A...173..375L} that the Mg\,\textsc{i} 12.32\,$\rm{\mu}$m line forms in the photosphere rather than the chromosphere.
The deviation of the highly excited level population of the Mg\,\textsc{i} atoms from the Bolzmann distribution makes this line transparent and turn into to emission in the upper photosphere.
{\citet{1995A&A...293..225B} and \citet{1995A&A...293..240B}} further synthesized the Mg\,\textsc{i} 12.32\,$\rm{\mu}$m line in active-region and {quiet-Sun} models, finding that its formation heights in the penumbra and the quiet Sun are similar. 
\citet{2021A&A...646A..79L} studied the radiation transfer process of the Mg\,\textsc{i} 12\,$\rm{\mu}$m pair line and derived a magnetic field calibration curve via wavelength integration of the Stokes profiles based on a one-dimensional solar atmospheric model. 
\citet{2024A&A...686A.278S} further tested the applicability and limitations of the wavelength-integrated method and the weak-field approximation for extracting the magnetic field based on a three-dimensional atmospheric model. 
\citet{2020ApJ...898..134H} investigated the evolution of the Mg\,\textsc{i} 12.32\,$\rm{\mu}$m line profile during flare heating using a flare atmospheric model, finding that heating initially reduces the line intensity before subsequently enhancing it. For polarization signals, flare heating decreases both the Zeeman splitting width and the Stokes~$V$ lobe amplitudes.

Given the potential of the Mg\,\textsc{i} 12.32 $\rm{\mu}$m line with regard to magnetic-field measurements, the Accurate Infrared Magnetic Field Measurements of the Sun (AIMS) Telescope has identified it as a primary scientific target and will provide more observations in the future \citep{2016ASPC..504..293D}.
In this paper, we attempt to address the underlying causes of the distinct behaviors of the Mg\,\textsc{i} 12.32 $\mu$m line in the quiet Sun and sunspot umbra, elucidating the relationship between its peak intensity and the physical parameters of the solar atmosphere. In Section~2, we describe the methodology and atmospheric models used in this study. Section~3 presents the synthesized Mg\,\textsc{i} 12.32 $\mu$m line profiles and response functions. In Section~4, we discuss the dominant physical parameters that influence the normalized peak intensity of the Mg\,\textsc{i} 12.32 $\mu$m line and their associated formation mechanisms. Our conclusions are summarized in Section~5.

\section{{Models and methods}}

\subsection{{Model atmosphere}}\label{ModelAtmo}
In this work, we utilized a snapshot from a three-dimensional radiation MHD simulation of a sunspot and the surrounding quiet {Sun} created with MURaM \citep{2005A&A...429..335V,2017ApJ...834...10R}.
This model contains a region with a horizontal area of 98.304$\times$49.152 Mm$^2$ and extends vertically for 49.152 Mm, covering atmospheric layers from the upper convection zone to the corona. 
The domain is resolved by a meshed of 1024$\times$512$\times$1536 grid points, corresponding to horizontal and vertical grid spacing of {96 km and 32 km}, respectively.
This model resembles an asymmetric sunspot pair and is evolved for more than ten solar hours. The analysis of the coronal dynamics of this model and a comparison with observations was shown by \citet{2024ApJ...973L...1L}.
{Figure}~\ref{fig1} shows the temperature of this model at 7.392 Mm above the bottom boundary of the model. 
In this work, we focused on the sunspot umbra and the surrounding quiet {Sun}, which is marked by the red rectangle in Fig.~\ref{fig1}. The former corresponding to a horizontal area of 5.28$\times$4.8 Mm$^2$, or 55$\times$50 in terms of grid points, and the latter corresponding to 6.72$\times$6.72 Mm$^2$, or 70$\times$70 in terms of grid points. Considering that the Mg\,\textsc{i} 12.32 $\rm{\mu}$m line is formed in the photosphere \citep{1987A&A...173..375L,1992A&A...253..567C}, we clipped the vertical height range from 4.8 to 11.2 Mm above the bottom boundary, which includes a part of the convection zone, the entire photosphere, and the chromosphere.

\begin{figure*}[h!]
\centering
\includegraphics[width=\hsize]{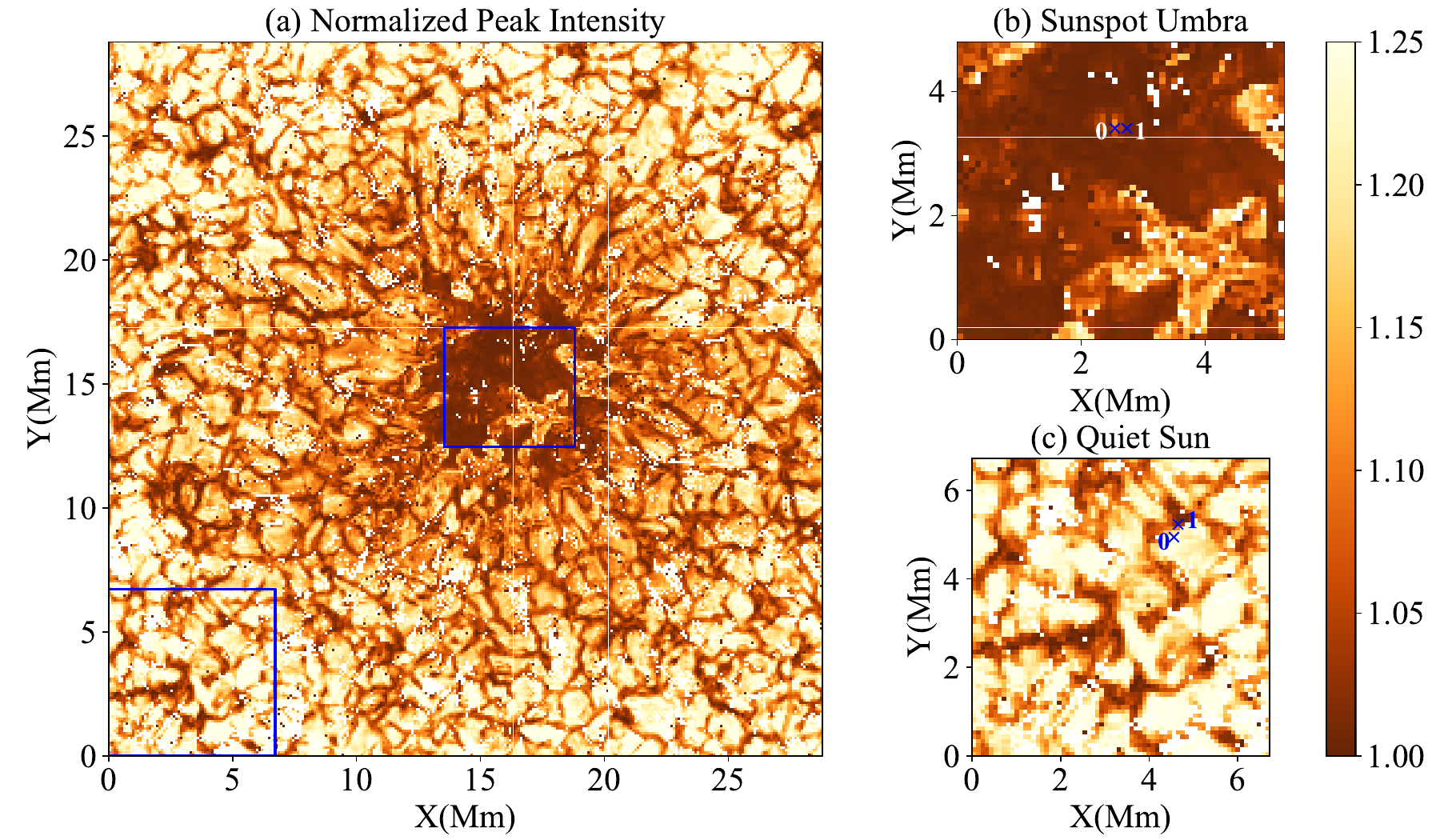}
\caption{
Panel (a): Normalized peak intensity of Mg\,\textsc{i} 12.32 $\rm{\mu}$m line calculated by RH 1.5D for MURaM model. Panel (b): Sunspot umbra. Panel (c): Quiet {Sun}. The blue crosses in panels (b) and (c) mark the pixels used for the analysis in Section 3.
Note that the white sprinkles indicate points where the radiative transfer calculation fails to converge {or crashes}.
}
\label{fig2}
\end{figure*}

\begin{figure*}[h!]
\centering
\includegraphics[width=\hsize]{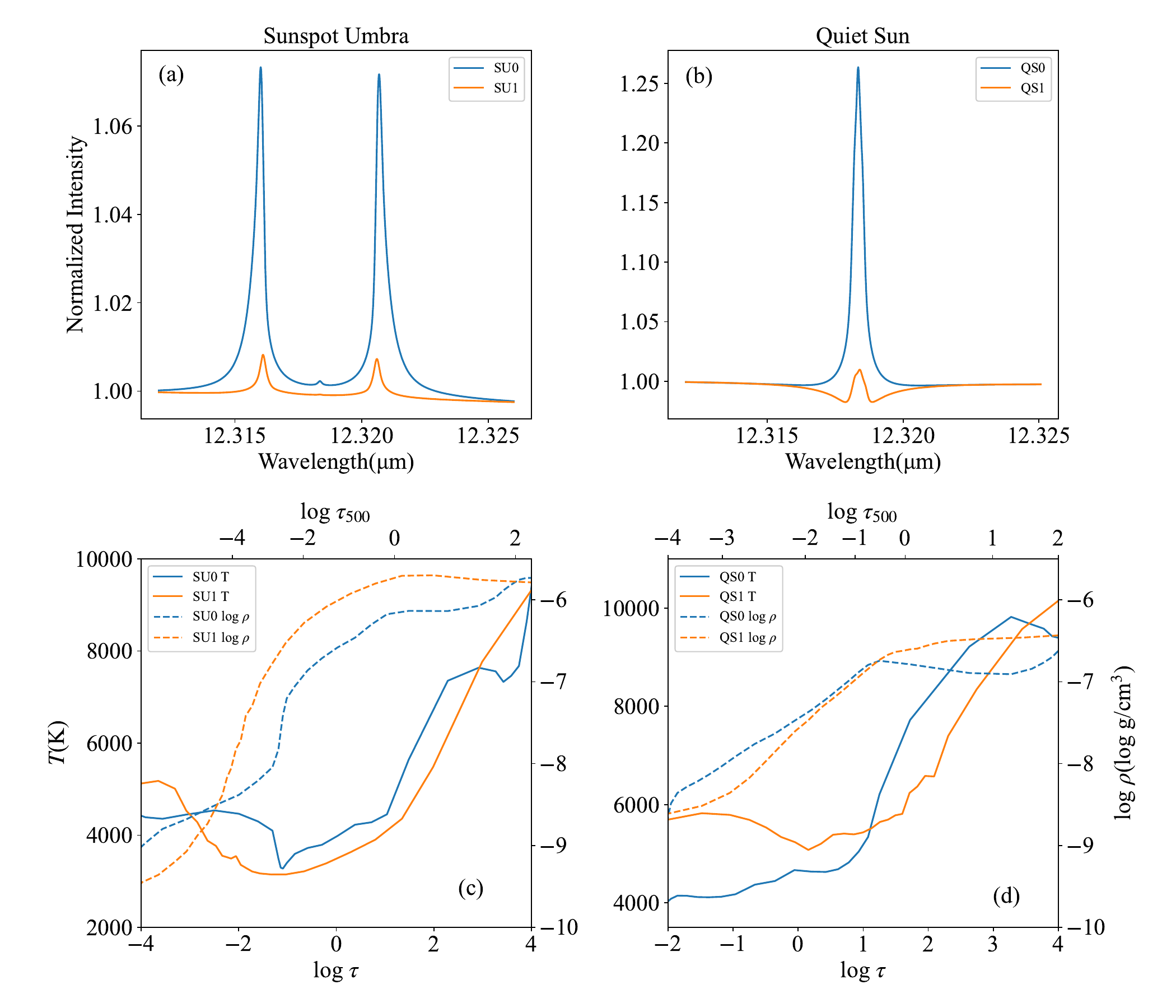}
\caption{
Upper panel: Normalized intensity profiles. Lower panel: {Temperature (left y-axis) and density (right y-axis) for the pixels in sunspot umbra (left panel) and quiet Sun (right panel) marked by blue crosses in {Fig.}~\ref{fig2}}. SU0: strong emission feature in sunspot umbra; SU1: weak emission feature in sunspot umbra; QS0: strong emission feature in granule; QS1: weak emission feature in intergranular lane.
Two optical depth scales are used in the lower panel, where optical depth, $\tau$, scales at the peak wavelength of the Mg\,\textsc{i} 12.32 $\rm{\mu}$m line intensity (lower x-axis), and $\tau_{500}$ is the continuum optical depth at $\lambda=500$ nm (upper x-axis). Note that the scale of $\tau_{500}$ is uneven.
}
\label{fig3}
\end{figure*}

\begin{figure*}[h!]
\centering
\includegraphics[width=\hsize]{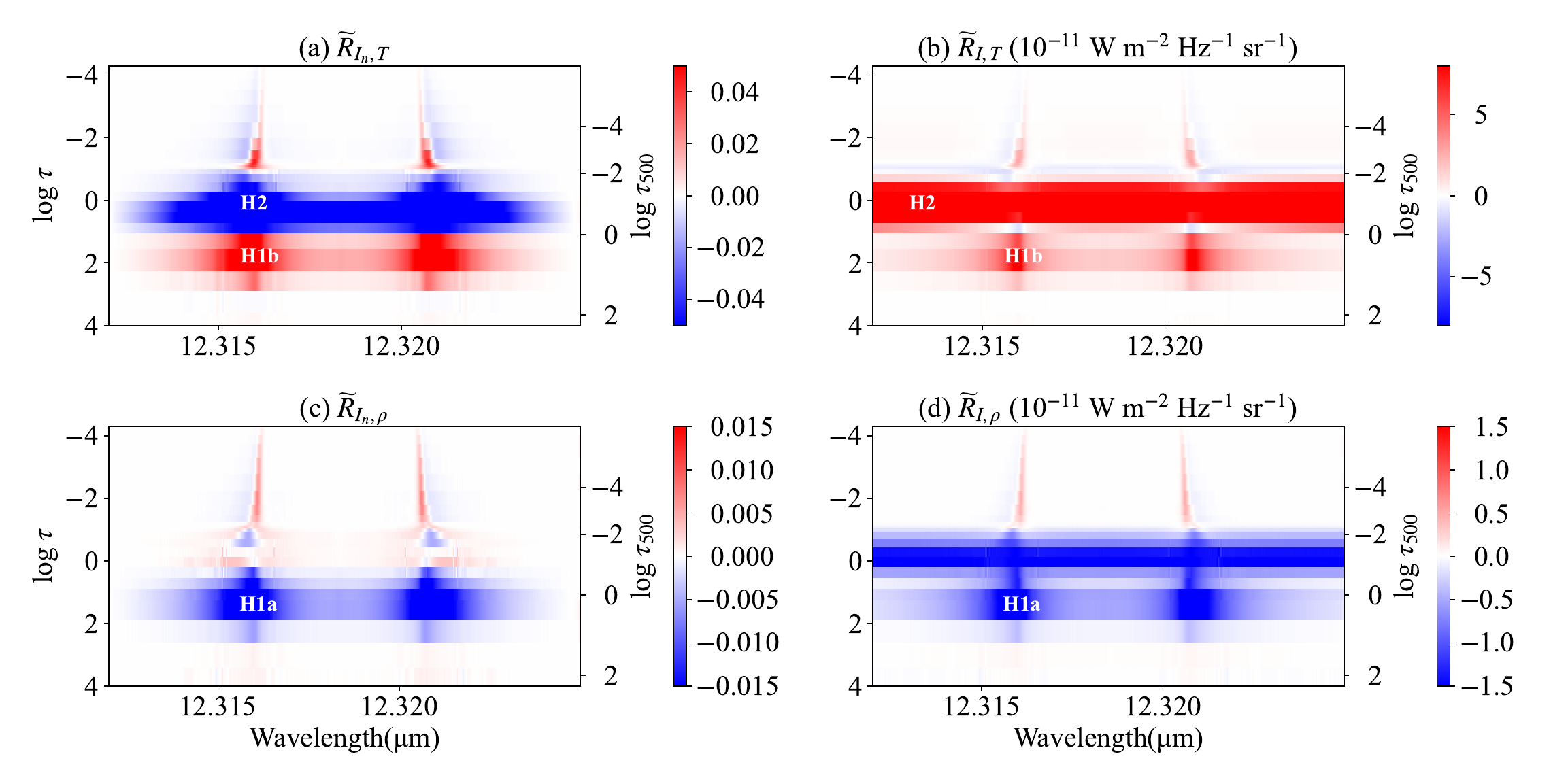}
\caption{Relative response functions of model SU0. {{Optical depth, $\tau$, scales at the peak wavelength of the Mg\,\textsc{i} 12.32 $\rm{\mu}$m line intensity, and $\tau_{500}$ is the continuum optical depth at $\lambda=500$ nm.} Note that the scale of $\tau_{500}$ is uneven.} Panel (a): Relative response function of normalized intensity to temperature, $\widetilde{R}_{I_{norm},T}$. Panel (b): Relative response function of intensity to temperature, $\widetilde{R}_{I,T}$. Panel (c): Relative response function of normalized intensity to density, $\widetilde{R}_{I_{norm},\rho}$. Panel (d): Relative response function of intensity to density, $\widetilde{R}_{I,\rho}$. The relevant definitions are given in Eq. (\ref{eq13}).
}
\label{fig4}
\end{figure*}

\begin{figure*}[h!]
\centering
\includegraphics[width=\hsize]{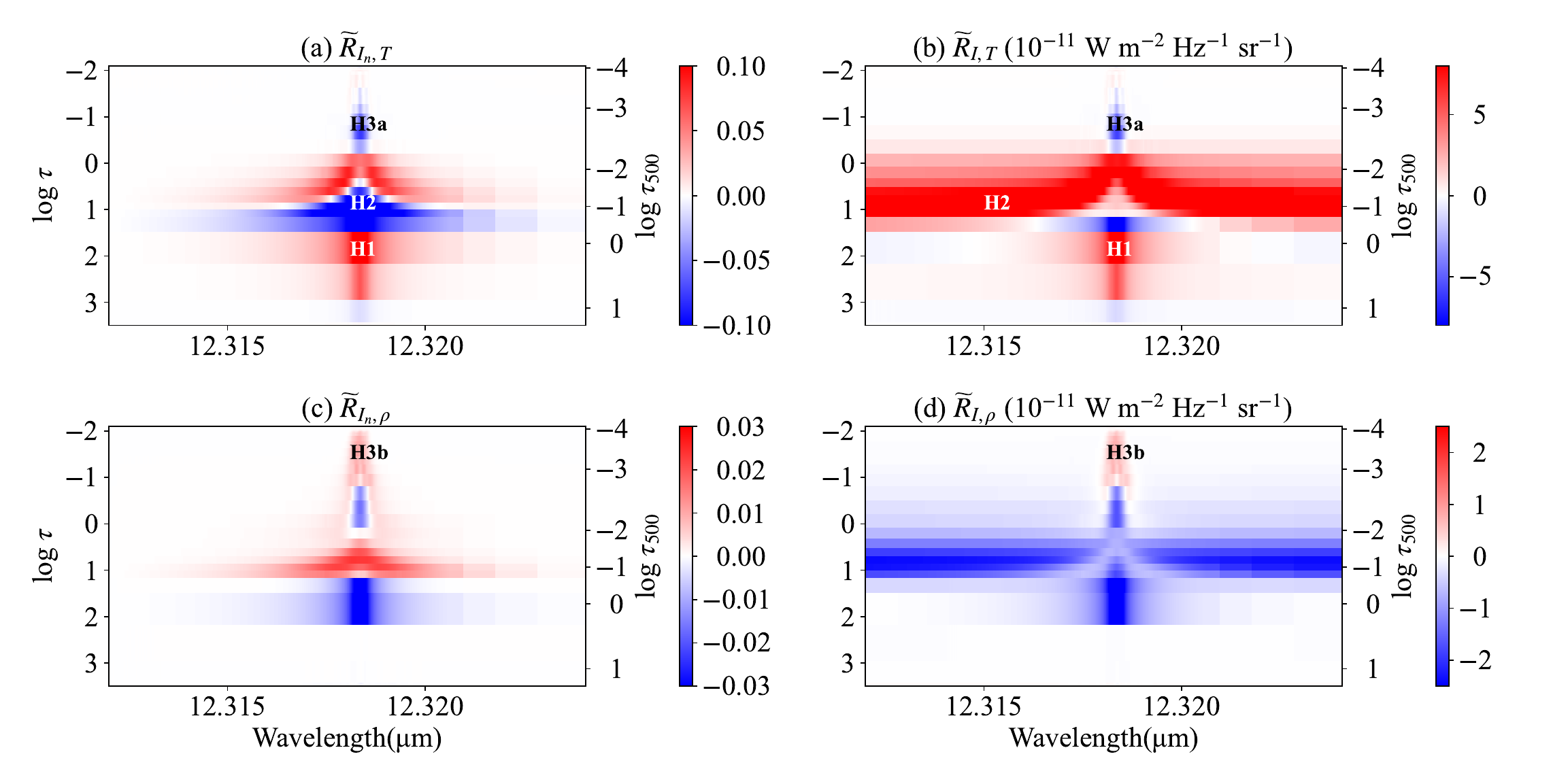}
\caption{Same as {Fig.}~\ref{fig4}, but for model QS0.}
\label{fig5}
\end{figure*}

\subsection{Radiative transfer code}\label{RTcode}
We used the Rybicki-Hummer (RH)\,1.5D code to solve the radiative transfer equation and synthesize Mg\,\textsc{i}\,12.32\,$\rm{\mu}$m line profiles \citep{2015A&A...574A...3P}. 
This code represents a parallelized improvement over the original \textsc{RH} code \citep{2001ApJ...557..389U}, which utilized the multilevel approximate lambda iteration formalism developed by \citet{1991A&A...245..171R,1992A&A...262..209R}.
The \textsc{RH\,1.5D} code incorporates NLTE effects, which are crucial for an accurate modeling of the Mg\,\textsc{i}\,12.32\,$\rm{\mu}$m line formation \citep{1992A&A...253..567C}. We adopted the \texttt{MgI\_66.atom} model atom from \citet{1992A&A...253..567C} and updated the values of the level energies based on NIST Atomic Spectra Database \citep{NIST_ASD} data. 
{We assumed complete frequency redistribution in our calculation, as the Mg\,\textsc{i}\,12.32\,$\rm{\mu}$m line is a transition between Rydberg states, and the partial frequency redistribution can be ignored.}
All calculations were performed assuming observations at the disk center (i.e., $\mu = 1$, where $\mu$ = cos$\theta$ and $\mathrm{\theta}$ is the heliocentric angle).

\subsection{Taylor expansion of the normalized peak intensity}\label{Taylor}

In this study, we proposed a method based on the Taylor expansion to identify the physical parameters at specific heights that have the greatest impact on the normalized peak intensity, and we believed this method can be generalized to study other characteristics of line profiles. 
The principle idea is that if the normalized peak intensity of a spectral line can be expressed as an analytical function of the physical parameters, then the difference in normalized peak intensity between two different atmospheric models can be approximated by the lower order terms of the Taylor series of this function.

The normalized intensity can be defined as $I/I_c$, where $I$ is the intensity and $I_c$ is the continuum. We denote the normalized peak intensity as $I_{norm}$, which is defined as the maximum of the normalized intensity of the Mg\,\textsc{i} 12.32 $\rm{\mu}$m line. In the case of complete Zeeman splitting, $I_{norm}$ is defined as the peak intensity of the most prominent Zeeman component.

The physical parameters of a model atmosphere can be represented as a single vector of $\bm{X}=[x_1\ \dots \ x_n]^{\rm{T}}$.
The response function, $R_{x_i}$, of $I_{norm}$ to the physical parameter, $x_i,$ can be defined as \citep{1994A&A...283..129R}
\begin{equation}
\label{eq1}
R_{x_i}=\frac{\partial I_{norm}}{\partial x_i}{.}
\end{equation}

For two different atmospheric models with the physical parameters $\bm{X}_0$ and $\bm{X}_1$, we denote their corresponding normalized peak intensities as $I_{norm}(\bm{X}_0)$ and $I_{norm}(\bm{X}_1)$, respectively. 
The difference in the physical parameters, $\bm{\Delta X}=[\Delta x_1\ \dots \ \Delta x_n]^{\rm{T}}$, is defined as $\bm{\Delta X}=\bm{X}_1-\bm{X}_0$. 
Assuming that $I_{norm}$ is an analytic function of the physical parameters $\bm{X}$, we can write the Taylor expansion of $I_{norm}$ centered at $\bm{X}_0$ as
\begin{equation}
\label{eq2}
\begin{aligned}
I_{norm}(\bm{X}_1) = & I_{norm}(\bm{X}_0)+\sum_{i=1}^n\frac{\partial I_{norm}(\bm{X}_0)}{\partial x_i}\Delta x_i \\
 + &\sum_{i=1}^n\sum_{j=1}^n\frac{1}{2}\frac{\partial^2 I_{norm}(\bm{X}_0)}{\partial x_i\partial x_j}\Delta x_i\Delta x_j+o(\lvert \bm{\Delta X}\rvert^2).
\end{aligned}
\end{equation}

Similarly, we can write the Taylor expansion of a response function, $R_{x_i}$, as
\begin{equation}
\label{eq3}
\begin{aligned}
R_{x_i}(\bm{X}_1)= & \frac{\partial I_{norm}(\bm{X}_1)}{\partial x_i} \\
= & R_{x_i}(\bm{X}_0)+\sum_{j=1}^n\frac{\partial^2 I_{norm}(\bm{X}_0)}{\partial x_i\partial x_j}\Delta x_j+o(\lvert \bm{\Delta X}\rvert),
\end{aligned}
\end{equation}
where $o(\lvert \bm{\Delta X}\rvert^2)$ and $o(\lvert \bm{\Delta X}\rvert)$ are the Peano's form of remainders.

We denote the average response function of the atmosphere model with physical parameters $\bm{X}_0$ and $\bm{X}_1$ as $\overline{R}_{x_i}$. Substituting Eq. (\ref{eq3}), we obtain
\begin{equation}
\label{eq4}
\begin{aligned}
\overline{R}_{x_i} = & \frac{R_{x_i}(\bm{X}_0)+R_{x_i}(\bm{X}_1)}{2}  \\
= & \frac{\partial I_{norm}(\bm{X}_0)}{\partial x_i}+\sum_{j=1}^n\frac{1}{2}\frac{\partial^2 I_{norm}(\bm{X}_0)}{\partial x_i\partial x_j}\Delta x_j+o(\lvert \bm{\Delta X}\rvert).
\end{aligned}
\end{equation}

Substituting Eq. (\ref{eq4}) to Eq. (\ref{eq2}),
\begin{equation}
\label{eq5}
I_{norm}(\bm{X}_1)-I_{norm}(\bm{X}_0)=\sum_{i=1}^n\overline{R}_{x_i}\Delta x_i+ o(\lvert \bm{\Delta X}\rvert^2).
\end{equation}

In Eq. (\ref{eq5}), if the higher order terms can be neglected, the difference of $I_{norm}$ can be decomposed into $n$ terms corresponding to the $n$ physical parameters. Each term $\overline{R}_{x_i}\Delta x_i$ includes the first-order term $\frac{\partial I_{norm}(\bm{X}_0)}{\partial x_i}\Delta x_i$, the second-order diagonal term $\frac{1}{2}\frac{\partial^2 I_{norm}(\bm{X}_0)}{\partial x_i^2}\Delta x_i^2$, and half of the corresponding second-order cross terms $\frac{\partial^2 I_{norm}(\bm{X}_0)}{\partial x_i\partial x_j}\Delta x_i\Delta x_j\ (i\neq j)$. Therefore, $\overline{R}_{x_i}\Delta x_i$ can quantify the contribution of the physical parameter, $x_i$, to the difference in $I_{norm}$ between two different atmospheric models. It should be noted that the condition for the validity of using the Taylor polynomial for function estimation is that the higher order terms, that is, the remainder, $o(\lvert \bm{\Delta X}\rvert^2)$ in Eq. (\ref{eq5}), is relatively small compared to the total difference ($I_{norm}(\bm{X}_1)-I_{norm}(\bm{X}_0)$) and can be neglected.

\subsection{Decomposition of response function}\label{RF}

The formation height of a spectral line in the solar atmosphere can be obtained from the analysis of its contribution function. The contribution function can be decomposed into the source function, the opacity, and the optical depth, and we can study the specific contribution of each term to the formation of the spectral line \citep{1997ApJ...481..500C,2018A&A...611A..62B}. 
Similarly, we can also decompose the response function in the same way, that is, into the source function, opacity, and optical depth, to study the contributions of different terms to the response function.

In this section, we refer to the definitions and derivations in Chapter 11 of \citet{1994ASSL..189.....S}.
Stokes $I$ can be expressed as a function of the contribution function, $C(z)$, or as a function of the source function, $S(z)$, opacity, $\chi (z)$, and optical depth, $\tau$:
\begin{equation}
\label{eq7}
I(0)=\int_0^{\infty}C(z)dz = \int_0^{\infty}S(z)\chi (z)e^{-\tau}dz;
\end{equation}
then, the contribution function can be written as
\begin{equation}
\label{eq9}
C(z)=S(z)\chi (z)e^{-\tau}.
\end{equation}

By taking the partial derivative of Eq. (\ref{eq7}) {to the physical parameter, $x_i$}, we can obtain the response function, $R_{x_i}$, of Stokes $I$ to {$x_i$}:
\begin{equation}
\label{eq10}
R_{x_i}=\frac{\partial I}{\partial x_i}=\int_0^{\infty}\frac{\partial C(z)}{\partial x_i}dz.
\end{equation}
Moreover, by computing the partial derivative of Eq.~\eqref{eq9}, the quantity $\frac{\partial C(z)}{\partial x_i}$ can be decomposed into three terms:
\begin{equation}
\label{eq11}
\frac{\partial C(z)}{\partial x_i}=\frac{\partial S(z)}{\partial x_i}\chi (z)e^{-\tau(z)}+\frac{\partial \chi (z)}{\partial x_i}S(z)e^{-\tau(z)}-\frac{\partial \tau(z)}{\partial x_i}S(z)\chi (z)e^{-\tau(z)},
\end{equation}
where the three terms correspond to the influence of changes in the source function, opacity, and optical depth on the response function, respectively. {We can change the independent variable to optical depth, $\tau$, and thus we can obtain the following:}

\begin{equation}
\label{eq11a}
\frac{\partial C(\tau)}{\partial x_i}=\frac{\partial S(\tau)}{\partial x_i}\chi (\tau)e^{-\tau}+\frac{\partial \chi (\tau)}{\partial x_i}S(\tau)e^{-\tau}-\frac{\partial \tau}{\partial x_i}S(\tau)\chi (\tau)e^{-\tau}.
\end{equation}

\section{Results}

\subsection{Synthetic profiles and normalized peak intensity}\label{profile}

Figure~\ref{fig2} shows the distribution of the normalized peak intensity of the Mg\,\textsc{i} 12.32\,$\rm{\mu}$m line in the sunspot umbra and quiet {Sun} (marked by the red rectangle in {Fig.}~\ref{fig1}), calculated using RH 1.5D based on the MURaM model atmosphere. 
Overall, the results are consistent with previous observations that the Mg\,\textsc{i} 12.32\,$\rm{\mu}$m line appears as an emission line in most parts of the quiet {Sun} {and the penumbra}, while it disappears in most parts of the sunspot umbra \citep{1983ApJ...269L..61B,1993ApJS...86..313H,2000ApJ...533.1035M}. Through high-spatial-resolution forward modeling, we also notice that (1) the Mg\,\textsc{i} 12.32\,$\rm{\mu}$m line shows strong emission in granules, while the emission is very weak in the intergranular lanes; (2) in the umbra, the emission of the Mg\,\textsc{i} 12.32\,$\rm{\mu}$m line disappears, except for some point features and block features where it is still present.

\begin{figure}[h!]
\centering
\includegraphics[width=\hsize]{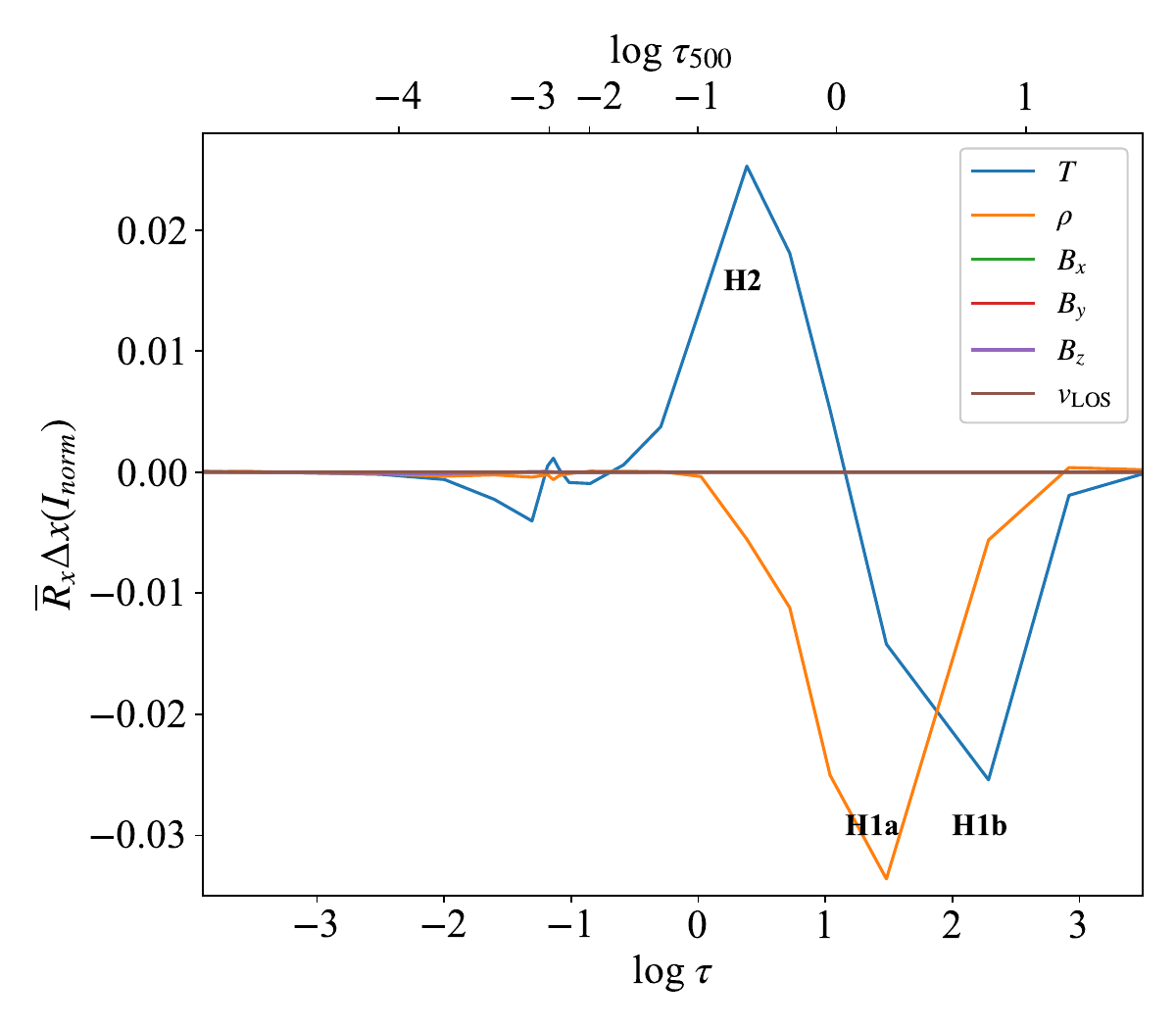}
\caption{$\overline{R}_{x_i}\Delta x_i$ for atmosphere model SU0 and SU1. Different colors represent different physical parameters, i.e., temperature, density{, the three magnetic-field components, and the LOS velocity}. $\tau$ is the optical depth at the peak wavelength of the Mg\,\textsc{i} 12.32 $\rm{\mu}$m line intensity, {and $\tau_{500}$ is the continuum optical depth at $\lambda=500$ nm for model SU0. Note that the scale of $\tau_{500}$ is uneven.}
}
\label{fig6}
\end{figure}

\begin{figure}[h!]
\centering
\includegraphics[width=\hsize]{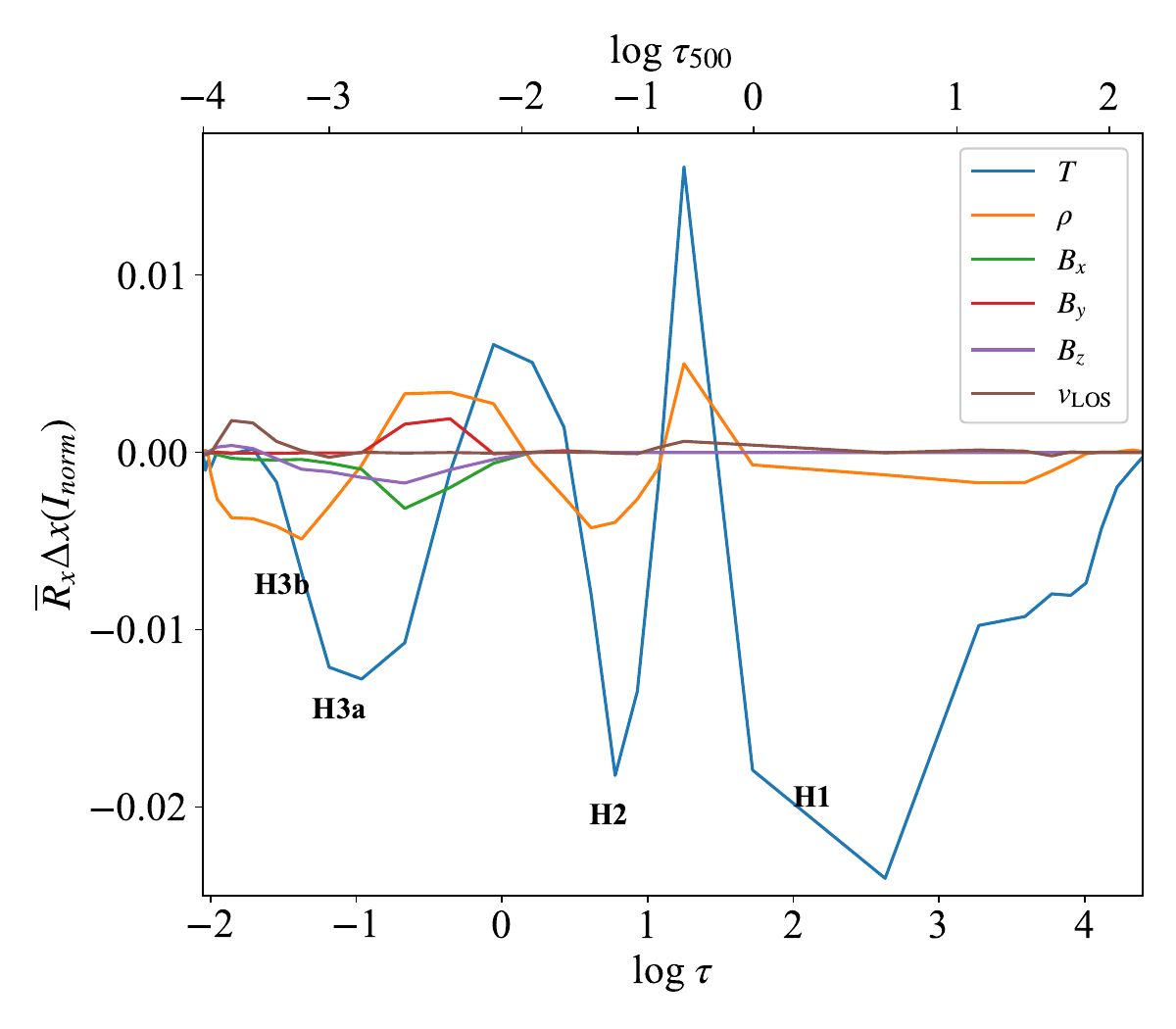}
\caption{
Same as {Fig.}~\ref{fig6}, but for model QS0 and QS1. {$\tau_{500}$ is the continuum optical depth at $\lambda=500$ nm for model QS0.}
}
\label{fig7}
\end{figure}

\begin{figure*}[h!]
\centering
\includegraphics[width=14cm]{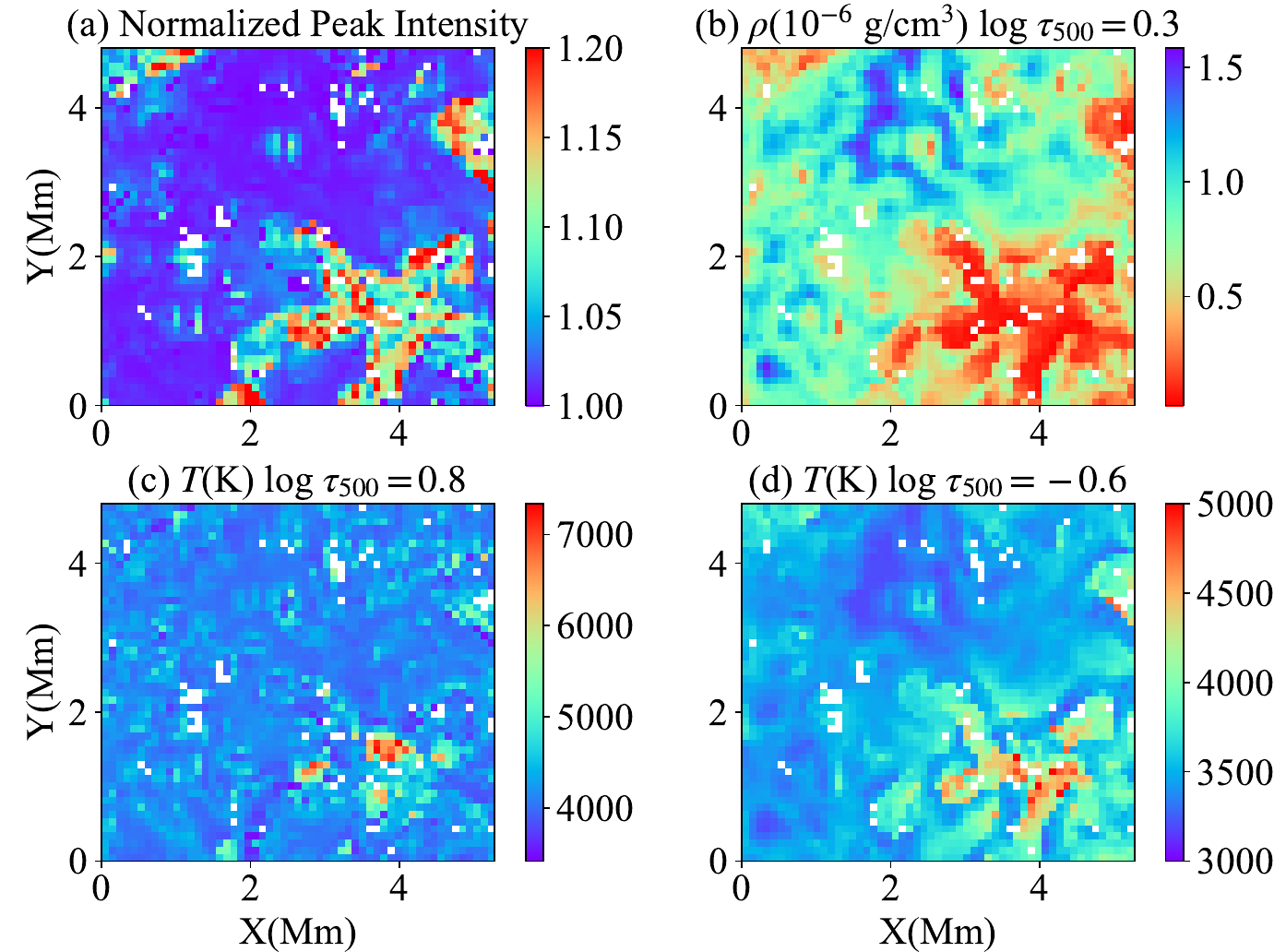}
\caption{Distribution maps in sunspot umbra. Panel (a): Normalized peak intensity; {panel (b): density at log $\tau_{500}=0.3$; panel (c): temperature at log $\tau_{500}=0.8$; panel (d): temperature at log $\tau_{500}=-0.6$.}}
\label{fig10}
\end{figure*}

\begin{figure*}[h!]
\centering
\includegraphics[width=\hsize]{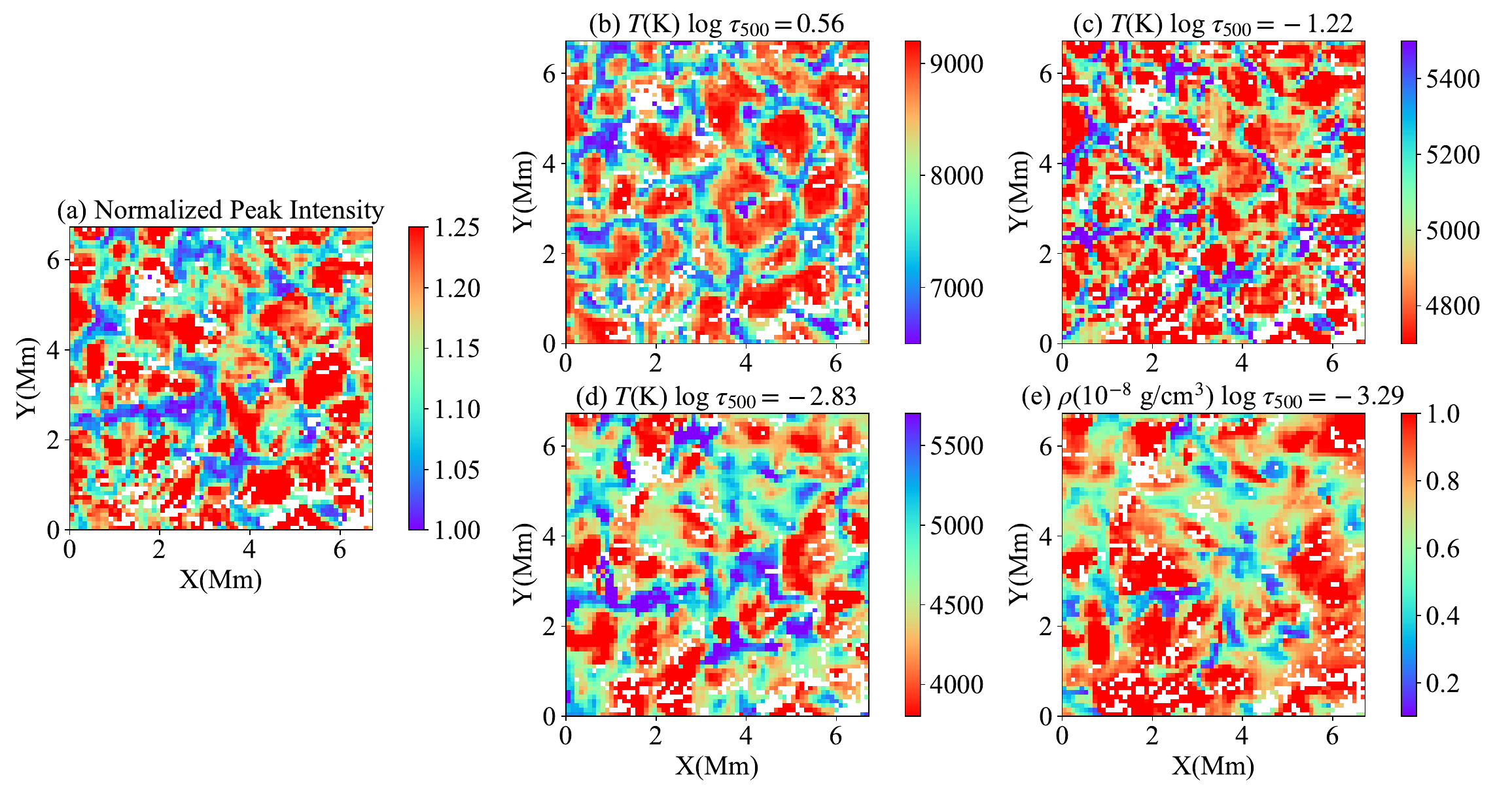}
\caption{Distribution maps in quiet {Sun}. Panel (a): normalized peak intensity; {panel (b): temperature at log $\tau_{500}=0.56$; panel (c): temperature at log $\tau_{500}=-1.22$; panel (d): temperature at log $\tau_{500}=-2.83$; panel (e): density at log $\tau_{500}=-3.29$.}}
\label{fig11}
\end{figure*}

In this work, we focused on the two features mentioned above, attempting to provide physical explanations for the distinct spectral characteristics observed in different regions. 
To this end, we selected two points in the umbra and granules, respectively, and
analyzed their response functions according to the method described in Section 2.3. 
These points are indicated by the blue crosses in panels (a) and (b) of {Fig.}~\ref{fig2}. 
The criteria for selecting these two points were that they should be spatially adjacent yet exhibit significantly different emission features; i.e., one with pronounced emission and the other with very weak emission. 
{Figure~\ref{fig3} displays the normalized intensity profiles, temperature, and density of the four points, with the left column representing the sunspot umbra and the right column representing the quiet {Sun}.}
As shown in the figure, in addition to the emission features, the synthesized spectral profiles also reproduced the observed features, including the absorption troughs and the complete line splitting in the sunspot umbra \citep{1983ApJ...269L..61B,1993ApJS...86..313H,2000ApJ...533.1035M}. The $\pi$ component is very weak in the umbra because of the nearly vertical magnetic field.

\begin{figure*}[h!]
\centering
\includegraphics[width=\hsize]{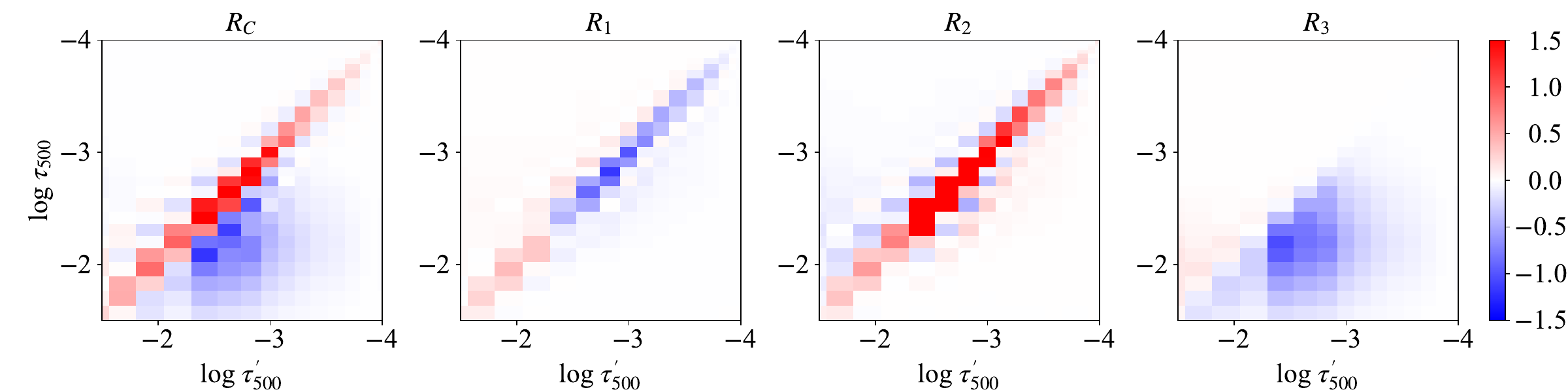}
\caption{$R_C,~R_1,~R_2,~R_3$ (in units of 10$^{-18}$ W~m$^{-3}~$Hz$^{-1}$~sr$^{-1}$~K$^{-1}$) as functions of {$\tau_{500}$ and $\tau^\prime_{500}$} for structure H3a.}
\label{fig12}
\end{figure*}

\begin{figure*}[h!]
\centering
\includegraphics[width=\hsize]{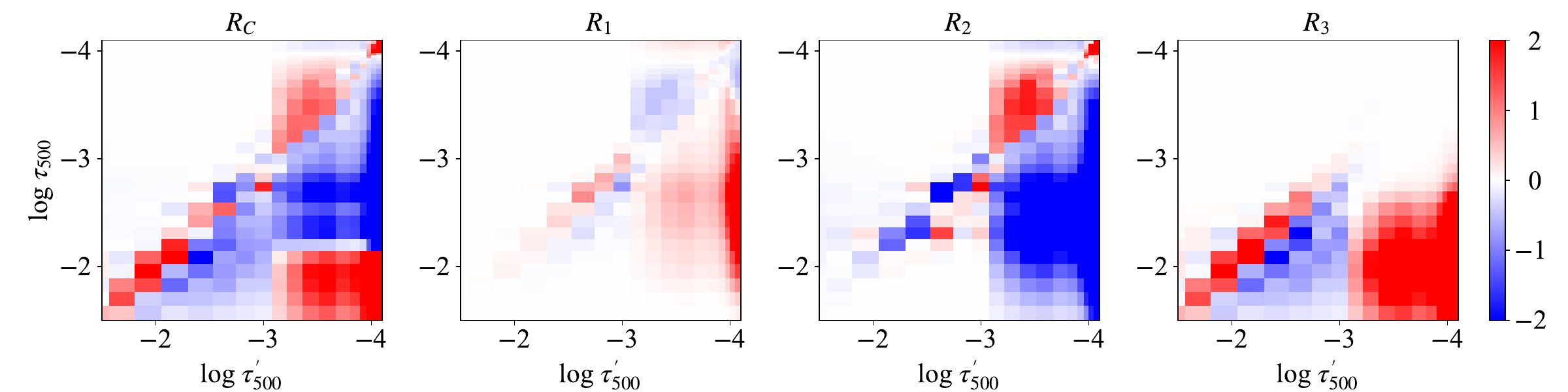}
\caption{$R_C,~R_1,~R_2,~R_3$ (in units of 10$^{-10}$ W~Hz$^{-1}$~sr$^{-1}$~kg$^{-1}$) as functions of {$\tau_{500}$ and $\tau^\prime_{500}$} for structure H3b.}
\label{fig13}
\end{figure*}

\subsection{Response function of intensity and normalized intensity}\label{IRF}

The response function provides information on the sensitivity of a spectral line to perturbations of the physical parameters of the atmospheric model \citep{1977A&A....56..111L}. When the response function exhibits a large positive or negative value at a certain layer and for a specific physical parameter, it indicates that the spectral line is highly sensitive to this parameter at that particular layer, and thus the spectral line can be utilized to diagnose this physical parameter there; if the response function has a low value or is close to zero, it implies that we cannot obtain reliable information about this parameter at that layer \citep{2016MNRAS.459.3363Q}. Therefore, the study of response functions is of significant guiding importance for future efforts to diagnose atmospheric physical parameters using the Mg \textsc{i} 12.32 $\rm{\mu}$m line.

Previous studies have revealed that the NLTE effect is essential in the formation of the Mg\,\textsc{i} 12.32 $\rm{\mu}$m line \citep[e.g.,][]{1992A&A...253..567C}. In this work, we calculated the response function by applying numerical perturbations to the physical parameters in the atmosphere model \citep[e.g.,][]{2004ApJ...603L.129S,2016MNRAS.459.3363Q}. 
As described in \citet{2016MNRAS.459.3363Q}, to calculate the response function of a physical parameter, $x$, at a height of $z$ for a model atmosphere, $M$, we created a new model $M^\prime$ that changed $x$ to $x+\Delta x$ at height $z$. 
It should be noted that temperature, $T$, density, $\rho$, and electron density, $n_e$, must satisfy the equation of state; therefore, if the physical parameter $x$ is temperature or density, we also need to change the electron density $n_e$ to $n_e+\Delta n_e$ in model $M^\prime$. 
Then, we synthesized the Mg\,\textsc{i} 12.32 $\rm{\mu}$m line profiles for model $M$ and $M^\prime$. In this way, we can obtain the response function of intensity $R_{I,x}$ and response function of normalized intensity $R_{I_{norm},x}$ by
\begin{equation}
\label{eq12}
R_{I,x}=\frac{I^{\prime}-I}{\Delta x},~~~~~~R_{I_{norm},x}=\frac{I^{\prime}_{norm}-I_{norm}}{\Delta x},
\end{equation}
where $I$ and $I_{norm}$ are intensity and normalized intensity calculated in model $M$. $I^\prime$ and $I_{norm}^\prime$ are the intensity and normalized intensity calculated in model $M^\prime$.

To better compare the relative sensitivity of the spectral line to different physical parameters at different heights, we calculated the relative response function $\widetilde{R}_{I,x}$ and $\widetilde{R}_{I_{norm},x}$ \citep{2016LRSP...13....4D}. The relative response function reflects the response of the spectrum to relative perturbations and is defined as
\begin{equation}
\label{eq13}
\widetilde{R}_{I,x}=R_{I,x}\cdot x,~~~~~~\widetilde{R}_{I_{norm},x}=R_{I_{norm},x}\cdot x.
\end{equation}

We calculated the response function of intensity $R_{I,x}$ and normalized intensity $R_{I_{norm},x}$ with respect to temperature, density, {magnetic field, and line-of-sight (LOS) velocity}, and then calculated $\overline{R}_{x_i}\Delta x_i$ according to the method described in Section 2.3. 
Figures~\ref{fig4} and \ref{fig5} show the relative response function, $\widetilde{R}_{I,x}$ and $\widetilde{R}_{I_{norm},x}$, to temperature and density, which are the two parameters that have the most significant impact on normalized peak intensity. More details are discussed in Section 4.1.

\section{Discussion}

\subsection{{The term $\overline{R}_{x_i}\Delta x_i$ and normalized peak intensity difference}}

By analyzing the response functions of two pairs of atmospheric models, (SU0, SU1) and (QS0, QS1), we found that the structures of the response functions for a given physical quantity are similar for each pair of atmosphere models. Based on the method described in Section \ref{Taylor}, we calculated $\overline{R}_{x_i}\Delta x_i$ for the two pairs of atmosphere models, which are marked by blue crosses in {Fig.} \ref{fig2}. The results are shown in {Figs.}~\ref{fig6} and~\ref{fig7}, respectively, where lines with different colors correspond to different physical parameters, namely temperature, density{, the three magnetic-field components, and the LOS velocity}. 
To make the structures of the response functions coincide, SU1 and QS1 are shifted by {$-$}0.06 Mm and {$-$}0.02 Mm, respectively. In the figures, a large positive or negative value of $\overline{R}_{x_i}\Delta x_i$ indicates that the corresponding physical parameter $x_i$ significantly contributes to the normalized peak intensity at a certain height.

As shown in {Figs.}~\ref{fig6} and~\ref{fig7}, the curves corresponding to temperature and density show relatively large positive or negative values as expected, indicating that temperature and density have the most significant influence on the normalized peak intensity in both the sunspot umbra and quiet {Sun}. It should be noted that $\tau$ is the optical depth at the peak wavelength of the Mg\,\textsc{i} 12.32 $\rm{\mu}$m line intensity{, and $\tau_{\rm{500}}$ is the continuum optical depth at $\lambda=500$ nm.} From {Figs.}~\ref{fig6} and~\ref{fig7}, we can draw the conclusions listed below.

\begin{enumerate}
    \item For models SU0 and SU1, density significantly affects the normalized line intensity within the range of log $\tau$ from 0 to 2.9 {or log $\tau_{\rm{500}}$ from $-$0.96 to 1.26} (denoted as H1a). The temperature also exhibits substantial responses in the ranges of log $\tau$ from 1.15 to 3.1 {or log $\tau_{\rm{500}}$ from 0.05 to 1.44} (H1b) and from {$-$}0.7 to 1.15 {or log $\tau_{\rm{500}}$ from $-$1.62 to 0.05} (H2). However, these responses have negative and positive contributions, respectively, leading to a partial offset of their effects.

    \item For models QS0 and QS1, the differences in normalized peak intensity can be attributed to three height ranges: 
    \begin{itemize}
        \item H1: the influence of temperature in the range of log $\tau$ from 1.5 to 3.0{, or log $\tau_{\rm{500}}$ from $-$0.21 to 0.91.}
        \item H2: the influence of temperature in the range of log $\tau$ from 0.45 to 1.1{, or log $\tau_{\rm{500}}$ from $-$1.65 to $-$0.76.}
        \item H3: the combined influence of temperature (H3a) and density (H3b) in the range of log $\tau$ from {$-$}2 to {$-$}0.3{, or log $\tau_{\rm{500}}$ from $-$3.86 to $-$2.38.} 
    \end{itemize}
\end{enumerate}

To verify the validity of the above analysis results, as we mentioned in Section~\ref{Taylor}, it is necessary for the remainder term {$o(\lvert \bm{\Delta X}\rvert^2)$} to be sufficiently small. For this purpose, we calculated the remainder terms $o(\lvert \bm{\Delta X}\rvert^2)$ for the SU and QS model pairs. 
{$I_{norm}(\bm{X}_1)-I_{norm}(\bm{X}_0)$ are {$-$}0.065 and {$-$}0.254 for the SU and QS model pairs, and the remainder terms are {2.8$\times 10^{-5}$} and {$-$0.066}, respectively.}
The results show that in the SU and QS models, the remainder terms account for less than 1\% and approximately {26\%} of the total difference $I_{norm}(\bm{X}_1)-I_{norm}(\bm{X}_0)$, {respectively.} Furthermore, by considering higher order terms in the Taylor series expansion, the remainder terms could be further reduced.
{However, high-order terms are more complex and challenging to relate to corresponding physical mechanisms.} 
Therefore, we consider the proportion of the remainder terms to be sufficiently small {for qualitative analysis}, and therefore $\overline{R}_{x_i}\Delta x_i$ can be used for qualitative analysis of the height and physical parameters that cause differences in the normalized peak intensities.

To analyze the applicability of the above conclusions, we compared the distribution of physical quantities {at the center of the structures displayed in {Figs.} \ref{fig6} and \ref{fig7}} with the distribution of normalized peak intensity. Figures \ref{fig10} and \ref{fig11} show the corresponding distribution images for the umbra and quiet {Sun} selected from {Fig.} \ref{fig2}, respectively. 
For the sunspot umbra region in {Fig.} \ref{fig10}, we compared the distribution maps of the density at {log $\tau_{500}=0.3$}, and the temperature at {log $\tau_{500}=0.8$ and $-$0.6}. Figure \ref{fig10} clearly reveals an anti-correlation between the density at {log $\tau_{500}=0.3$} and the normalized peak intensity. {From panels (c) and (d), we hardly observe any correlation with the temperature distributions at log $\tau_{500}=0.8$ and $-$0.6.} 
Overall, the density $\rho$ at {log $\tau_{500}=0.3$} is the primary factor affecting the normalized peak intensity in the sunspot umbra region. This is consistent with the findings in {Fig.} \ref{fig6}. 
For the quiet {Sun} in {Fig.} \ref{fig11}, the distribution maps of the temperature at {log $\tau_{500}=0.56$, $-1.22,$ and $-2.83$}, and the density at {log $\tau_{500}=-3.29,$} show that all four heights have certain relationships with the normalized peak intensity in the quiet {Sun}, indicating that the factors influencing the relative intensity of the Mg\,\textsc{i} 12.32 $\rm{\mu}$m line in the quiet {Sun} are more complex than in the umbra region, as demonstrated in {Fig.} \ref{fig7}.

\subsection{Formation mechanisms}

\citet{1992A&A...253..567C} discussed the formation mechanism of the Mg\,\textsc{i} 12.32 $\mathrm{\mu m}$ spectral line.
The emission of the Mg\,\textsc{i} 12.32 $\mathrm{\mu m}$ line, as explained in the paper, is a consequence of population departures of the upper and lower levels of the 12.32 $\mathrm{\mu m}$ line, which is caused by the population depletion and replenishment from the Mg\,\textsc{ii} reservoir in a NLTE process. The depletion occurs primarily due to photon losses in lines with excitation energies around 6--7 eV that become optically thin in the photosphere. 
In Section 4.1, we discussed the relationship between the normalized peak intensity and certain physical parameters at specific heights. The purpose of this section is to attempt to explain how these parameters affect the line intensity. Since we used the normalized peak intensity as our research subject in this work, both factors affecting the continuum and the Mg\,\textsc{i} 12.32 $\mathrm{\mu m}$ line influence the results.

Based on the analysis of the response functions in {Figs.}~\ref{fig4} and~\ref{fig5}, 
the H1 heights in both SU and QS models lie below the formation height of the Mg\,\textsc{i} 12.32 $\rm{\mu}$m line. We propose that H1 height influence the spectral line with the mechanism discussed by \citet{1992A&A...253..567C}, where photon losses in spectral lines with excitation energies of approximately 6--7 eV lead to deviations in energy populations, thereby affecting the line intensity.
The H2 height in the QS model is located at the continuum formation height, affecting the normalized peak intensity through its contribution to the continuum. 
The H3 height in the QS model corresponds to the formation height of the Mg\,\textsc{i} 12.32 $\rm{\mu m}$ line. Here, we attempt to analyze the influence of the H3 height (H3a of the temperature and H3b of the density) on the line intensity using the decomposition method described in Section~2.3. We denote the physical parameters at {optical depth $\tau^\prime$} as
\begin{equation}
\label{eq14}
\begin{aligned}
R_1&=\frac{\partial S(\tau)}{\partial x(\tau^\prime)}\chi (\tau)e^{-\tau}; \\
R_2&=\frac{\partial \chi (\tau)}{\partial x(\tau^\prime)}S(\tau)e^{-\tau}; \\
R_3&=-\frac{\partial \tau}{\partial x(\tau^\prime)}S(\tau)\chi (\tau)e^{-\tau},
\end{aligned}
\end{equation}
where $R_1$, $R_2$, and $R_3$ correspond to the three terms on the right side of Equation (\ref{eq11a}), respectively, satisfying $R_C = R_1 + R_2 + R_3$, and $R_C$ is the response function of the contribution function. 
Figures \ref{fig12} and \ref{fig13} show $R_C$, $R_1$, $R_2$, and $R_3$ for the H3a and H3b structures corresponding to temperature and density, respectively. {$\tau$ and $\tau^\prime$ are converted to the continuum optical depth at $\lambda=500$ nm, $\tau_{500}$ and $\tau^\prime_{500}$, in Figs. \ref{fig12} and \ref{fig13}.}

From Figs.~\ref{fig5} and \ref{fig7}, it can be seen that the H3a height corresponds to a negative response function. In Fig.~\ref{fig12}, the main contributing factor to the negative response function is $R_3$ in the region of {$\tau_{500} > \tau^\prime_{500}$}, which represents the influence of optical depth on the response function. As shown by $R_2$, the increase in opacity attenuates the light intensity during propagation. Consequently, the intensity decreases, resulting in a negative response function. Furthermore, $R_1$ and $R_2$ indicate that the source function and opacity are minor contributing factors. For the H3a structure, $R_1$ and $R_2$ only show responses on the {$\tau_{500} = \tau^\prime_{500}$} line, suggesting that changes in the source function and opacity are independent of the nonlocal temperature.

Figure~\ref{fig13} indicates that the interpretation of the H3b structure is more complicated.
The variation in $R_2$ shows that the opacity increases at {$\tau_{500} = \tau^\prime_{500}$} and decreases at {$\tau_{500} > \tau^\prime_{500}$}. The response of $R_2$ in the {$\tau_{500} > \tau^\prime_{500}$} region demonstrates the nonlocal effect of density variations, meaning that changes in density can affect the opacity in the lower atmospheric layers. 
Subsequently, these changes in opacity enhance the light intensity during light propagation; therefore, $R_3$ shows a positive response when {$\tau_{500} > \tau^\prime_{500}$}. 

\section{Conclusions}

We synthesized the Mg\,\textsc{i}\, 12.32 $\rm{\mu}$m line using the RH 1.5D radiative transfer code for a three-dimensional radiation magnetohydrodynamic model of a sunspot. We investigated the distribution of normalized peak intensities of the Mg\,\textsc{i}\, 12.32 $\rm{\mu}$m line in different regions, namely {the umbra and quiet {-Sun} regions.} The key features are summarized below.

\begin{enumerate}
    \item In the quiet Sun, the Mg\,\textsc{i}\, 12.32 $\rm{\mu}$m line exhibits strong emission in the granules, while showing very weak emission in the intergranular lanes.
    \item {In the penumbra, the Mg\,\textsc{i}\, 12.32 $\rm{\mu}$m line exhibits strong emission. }
    \item In the umbral region, the Mg\,\textsc{i}\, 12.32 $\rm{\mu}$m line shows no obvious emission in most areas, except for a few points where emission is visible.
\end{enumerate}

We further investigated the causes of different emission features in the quiet-Sun and umbral regions separately. By calculating the $\overline{R}_{x_i}\Delta x_i$ parameters for two pairs of atmospheric models [SU0, SU1], and [QS0, QS1], where $\overline{R}_{x_i}$ is the mean response function and $x_i$ represents selected physical parameters, we identified the physical parameters responsible for the differences in normalized peak intensities. Our conclusions are as follows.

\begin{enumerate}
    \item In the sunspot umbra, the density in the region with log $\tau$ ranging from 0 to 2.9{, or log $\tau_{\rm{500}}$ from $-$0.96 to 1.26} (H1a), is the main factor causing the difference in normalized peak intensities between SU0 and SU1.
    \item In the quiet-Sun region, the difference in normalized peak intensities between QS0 and QS1 can be attributed to the temperature {in the range of log $\tau$ from 1.5 to 3.0, or log $\tau_{\rm{500}}$ from $-$0.21 to 0.91} (H1), and {log $\tau$ from 0.45 to 1.1, or log $\tau_{\rm{500}}$ from $-$1.65 to $-$0.76} (H2), as well as the combined effects of temperature and density {in the range of log $\tau$ from $-$2 to $-$0.3, or log $\tau_{\rm{500}}$ from $-$3.86 to $-$2.38} (H3).
\end{enumerate}

Furthermore, we discussed the possible physical mechanisms by which these heights and physical parameters influence the emission intensity of the 12.32 $\rm{\mu}$m spectral lines. 
\begin{enumerate}
\item H1 corresponds to the mechanism that photons with 6-7 eV excitation energy, which becomes optically thin in the photosphere, can drive high Rydberg levels populations of Mg\,\textsc{i}, thus contributing to the intensity of the Mg\,\textsc{i} 12.32 $\rm{\mu}$m line.

\item H2 is located at the formation height of the continuum. The physical parameters in H2 affect the continuum intensity, $I_{c}$, thus influencing the normalized peak intensity.

\item The primary reason for the negative response function of the temperature at the height of H3a is that an increase in temperature leads to higher opacity, thereby attenuating the radiation from deeper layers during transmission.
The positive response function of density at the height of H3b is influenced by multiple factors, including variations in opacity and the resulting enhancement of radiation originating from the lower layers of the atmosphere.
\end{enumerate}

In summary, this work investigated the emission characteristics of the Mg\,\textsc{i} 12.32 $\rm{\mu}$m spectral line in different solar regions. Through the analysis of response functions, we identified the key physical quantities and mechanisms that affect the relative intensity of the spectral line. This study will provide an important foundation for future research on the inversion of solar atmospheric parameters based on the Mg\,\textsc{i} 12.32 $\rm{\mu}$m spectral line.

\begin{acknowledgements}
This work is supported by National Key R\&D Program of China 2022YFF0503803, the National Natural Science Foundation of China grants 12373058, 12422308 and 12373054, the Specialized Research Fund for State Key Laboratory of Solar Activity and Space Weather and the Strategic Priority Research Program of the Chinese Academy of Sciences, grants No. XDB0560000. WL thank Han Uitenbroek and Tiago Pereira for their help with the RH and RH 1.5D code. {We would also like to thank the anonymous referee for her or his useful comments that helped improve the original manuscript.}
\end{acknowledgements}

\bibliographystyle{aa} 
\bibliography{aa_example}

\end{document}